# TITLE PAGE

## Title: ChatGPT Needs SPADE (Sustainability, PrivAcy, Digital divide, and Ethics) Evaluation: A Review.

First Author

Name: Sunder Ali Khowaja (Corresponding Author)

Affiliation: Faculty of Computing, Digital and Data, Technological University Dublin, Ireland.

Email: sunderali.khowaja@tudublin.ie

Second Author

Name: Parus Khuwaja

Affiliation: Institute of Business Administration, University of Sindh, Jamshoro, Pakistan.

Email: parus.khuwaja@usindh.edu.pk

Third Author

Name: Kapal Dev

Affiliation: Department of Computer Science, Munster Technological University, Cork, Ireland.

Email: kapal.dev@ieee.org

Fourth Author

Name: Weizheng Wang

Affiliation: City University of Hong Kong, Hong Kong SAR, China

Email: weizheng.wang@ieee.org

Fifth Author

Name: Lewis Nkenyereye

Affiliation: Sejong University, Seoul, South Korea

Email: nkenyele@sejong.ac.kr

# ChatGPT Needs SPADE (Sustainability, PrivAcy, Digital divide, and Ethics) Evaluation: A Review.

Sunder Ali Khowaja[1*], Parus Khuwaja[2], Kapal Dev[3], Weizheng Wang[4], and Lewis Nkenyereye[5]

[1] Faculty of Computing Digital and Data, Technological University Dublin, Ireland

[2] Institute of Business Administration, University of Sindh, Pakistan

[3] Nimbus Research Centre, Munster Technological University, Ireland

[4] City University of Hong Kong, Hong Kong SAR, China

[5] Sejong University, Seoul, South Korea

*Corresponding Author

sunderali.khowaja@tudublin.ie , parus.khuwaja@usindh.edu.pk, kapal.dev@ieee.org, weizheng.wang@ieee.org, nkenyele@sejong.ac.kr

## Abstract

ChatGPT is another large language model (LLM) vastly available for the consumers on their devices but due to its performance and ability to converse effectively, it has gained a huge popularity amongst research as well as industrial community. Recently, many studies have been published to show the effectiveness, efficiency, integration, and sentiments of chatGPT and other LLMs. In contrast, this study focuses on the important aspects that are mostly overlooked, i.e. sustainability, privacy, digital divide, and ethics and suggests that not only chatGPT but every subsequent entry in the category of conversational bots should undergo Sustainability, PrivAcy, Digital divide, and Ethics (SPADE) evaluation. This paper discusses in detail the issues and concerns raised over chatGPT in line with aforementioned characteristics. We also discuss the recent EU AI Act briefly in accordance with the SPADE evaluation. We support our hypothesis by some preliminary data collection and visualizations along with hypothesized facts. We also suggest mitigations and recommendations for each of the concerns. Furthermore, we also suggest some policies and recommendations for EU AI policy act concerning ethics, digital divide, and sustainability

## 1. Introduction

Technology has advanced manifold since the first statistical model designed for language understanding. Since the inception of deep learning techniques and availability of large-scale data, language models have seen drastic improvement in terms of language understanding tasks while surpassing human-level performance at times. Over the years, researchers have developed a keen interest in implementing and improving large language models (LLMs) using variants of deep learning architectures [1]. The LLMs are trained on large-scale textual datasets and learn to model linguistic characteristics for generating sensible, coherent, and conversational responses to natural language queries. The LLMs are also considered for text generative systems that could help in creating responses and generating novel texts while providing customized text-based prompts. These generative systems have been used extensively for language translation, question answering systems and chatbot designs, respectively. Although many deep learning techniques have contributed to the design of LLMs but most of the success has been attributed to Transformer architecture that was introduced in [2]. The study introduced the sequence processing with self-attention mechanism that replaced the conventional network architectures including recurrent neural networks (RNNs), gated recurrent units (GRUs), and long-short term memory (LSTM) networks. Due to the capability of self-attention, the model focuses on selective parts of the sequence, which helps the network to learn contextual linguistic information, hence, is better in generating customized output sequences. Transformer networks have been extensively used in applications concerning question answering systems, machine translation, language modeling, and vision related tasks. Furthermore, the self-attention mechanism helps in modelling long-range dependencies that is helpful in generating long texts instead of short answers. Considering the current and most powerful LLMs are based on Transformer architecture at their core, there is no denying that Transformer architectures have contributed to the extended success of the LLMs in recent years.

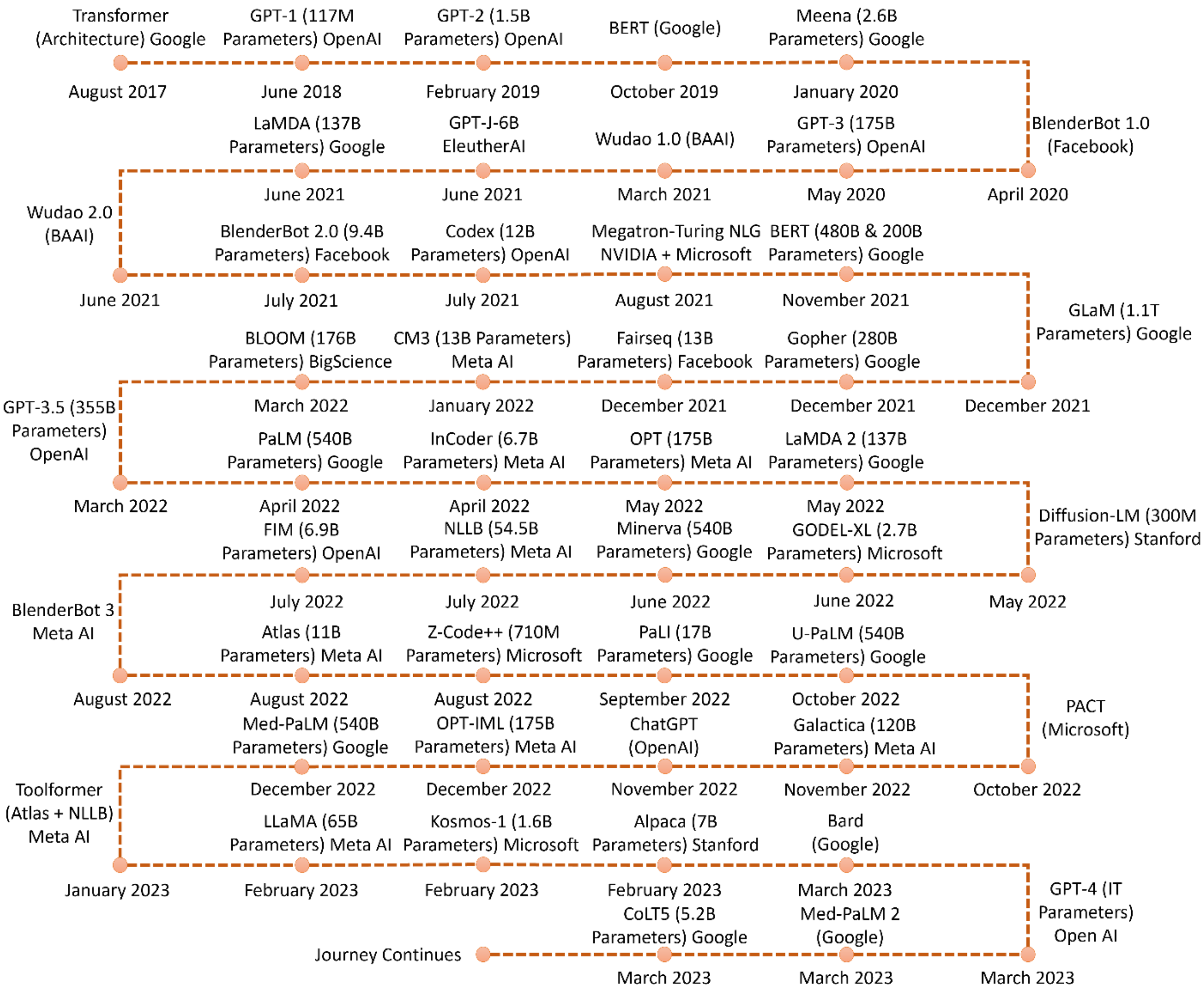

Figure 1 A Brief Timeline of Large Language Models

With the success of LLMs, a growing interest amongst researchers from within and outside of the computer science community has also been observed for artificial intelligence generated content (AIGC). The interest has been increased due to the launch of powerful LLMs from various companies including Google, OpenAI, Microsoft, and Huggingfaces. Some of them are limited to a single modality such as ChatGPT[1], while others take into account multi-modal data such as GPT-4 [3]. AIGC refers to the content generation using advanced generative AI (GAI) techniques in an automated way contrasting to the human invasive approach. For instance, ChatGPT designed by OpenAI understands inputs provided by humans and responds through textual modality in a meaningful manner. Until the release of GPT-4, chatGPT was considered to be the most powerful conversational bot that has ever been released to the public. On the other hand, Dall-E2, also designed by OpenAI undertakes textual description from the humans and generates high quality images. The release of a few LLMs in chronological order is shown in Figure 1. Although many of the LLMs have been included in Figure 1, but it should be noted that the list is not complete in order to be comprehensive. There are many competitors such as DeepMind, Amazon, EleutherAI, BigScience, Aleph Alpha, Huawei, Tsinghua, Together, Baidu, and many others that are not included in the given timeline.

[1] https://openai.com/blog/chatgpt

The content generation with AIGC utilizes GAI algorithms along with human instructions to guide and teach the model for task completion and satisfy the instruction. Mainly two steps are considered for such content generation: the first is related to the understanding of human intent from provided instructions and the second is to generate the content based on the identified intention. Although, carrying out the above two steps are similar in most of the studies (from basic methodology point of view), the advancements are observed due to the increased computational resources, larger model architectures, and availability of large-scale datasets. An example of a transition from GPT-2 to GPT-3 can better illustrate the aforementioned reasoning. The main framework of both the GPTs are the same, however, both of them differ in foundation model size, i.e. 1.5 billion and 175 billion, and the pre-training data, i.e. WebText [4] and CommonCrawl [5], respectively. CommonCrawl is 15x larger than WebText. The results are quite evident as GPT-3 extracts human intentions in a better way while generalizing well to human instructions in comparison to GPT-2. Currently, the number of parameters for GPT-4 have not been released officially, but it is safe to assume that the number of parameters will be higher than its predecessor.

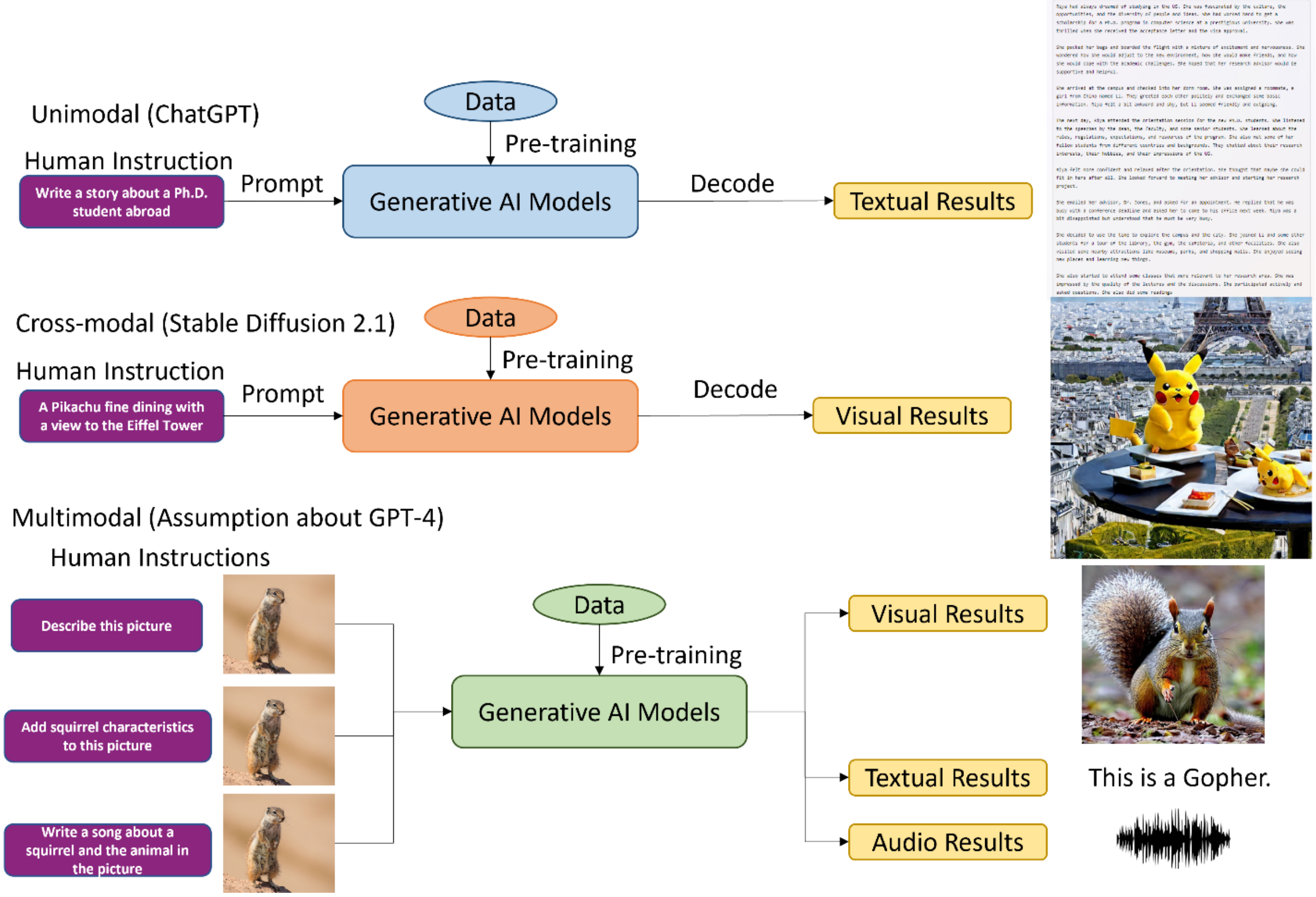

Figure 2 A comparison between unimodal, cross-modal, and multimodal Generative AI models. For unimodal and cross-modal, the results are generated using chatGPT and stable diffusion 2.1. For the multimodal, the results are assumptions as the access to GPT-4 is still limited.

Another dimension apart from computational resources and data availability is the design of algorithms that improve the appropriateness, responsiveness and consumer-level deployment of the GAI frameworks. For instance, accuracy and reliability of generated responses in accordance to human queries using chatGPT is attributed to reinforcement learning from human feedback (RLHF) [6–8]. RLHF is the method that allows chatGPT to generate long dialogues and converse better with humans. Similarly for the field of computer vision, Stability.AI proposed stable diffusion to generate high quality images [9] based on human intentions exhibited by text prompts. Stable diffusion achieves better trade-off in terms of exploitation and exploration, thus generating high-quality images that are both similar to the training data and diverse enough for the humans to perceive it as a unique generation. GPT-4 combines both of the characteristics by undertaking textual as well as image input for generating the output. chatGPT was a unimodal GAI, stable diffusion was a cross-modal GAI, while GPT-4 is a multimodal GAI, respectively. An illustration distinguishing between unimodal, cross-modal, and multimodal GAIs are shown in Figure 2.

The combination of unimodal and cross-modal GAIs have resulted in various startups, basis of new research works, and industrial implications in recent times. The implications can be found in areas but not limited to education [10], advertising [10,11], and art [11,12]. It is assumed that the GPT-4 will extend its footprint to even further domains at a significant pace.

Considering the popularity and chatGPT user subscription, it is therefore, important to not only know but also evaluate these GAI algorithms such as GPT-4 in terms of sustainability, privacy, digital divide, and ethics (SPADE). The rationale for conducting this study is to provide the basis for SPADE evaluation for LLMs and generative AI systems such as ChatGPT. Although EU AI Act is recently released but many of the details are still not provided, such as the process of evaluating an AI system, the process for limiting the energy resources, the process for levelling the field in developed and developing countries, and the process to preserve the privacy of the users as well as existing copyrighted material. In this regard, this work discusses some of the concerns related to the aforementioned characteristics, discusses them, and provides a basis for policy changes and review for honor code, respectively.

The rest of the paper is structured as follows. Section 2 highlights the selection criteria. Section 3 provides details regarding the sustainability issue. Section 4 provides details concerning privacy issues. Section 5 presents details regarding the digital divide issue. Section 6 provides details concerning ethical issues and Section 7 discusses the EU AI Act in relation to the generative AI systems. We lastly provide lessons learned in Section 8 and conclude this study in Section 9.

## 2. Selection Criteria

It is always important to lay out the selection criteria for such study. However, it's not the same with the conventional existing studies as it highlights some of the important aspects concerned with large language models (LLMs), specifically ChatGPT. Furthermore, to the best of our knowledge, there is no one study that highlights the collective problems of LLMs with respect to sustainability, privacy, digital divide and ethical considerations. In addition, all of the review studies related to LLMs and ChatGPT are in constant flux, therefore, we cannot adopt the conventional selection criteria for selecting the studies. It should also be noted that every iteration of the proposed work is reviewed, and more related works are added, specifically with the concerns listed above.

Considering the aforementioned concerns, as we are dealing with multiple concerns regarding LLMs and ChatGPT while dealing with multiple iterations of this work, we would like to layout the steps that we followed for the compilation of the proposed study.

- For each of the concerns, we used keywords that highlight the specific LLM issue, such as Sustainability, Privacy, Digital Divide, and Ethics.
- For Sustainability, we further used keywords such as energy consumption, carbon offset, LLM Training Cost, and LLM Testing Cost.
- For Privacy, we further used keywords such as data privacy, copyrighted text in LLMs, user privacy, and data ownership.
- For Digital Divide, we further used keywords LLM accessibility, Third World Countries, Internet Accessibility, and Human development index.
- For Ethics, we further used keywords such as Regulation Acts, Ethical AI, Ethical LLM, and EU AI Act.

Following the search criteria, we came across several articles, papers, and interviews that can be categorized as verified and unverified. Among the works that were returned in the search, we only considered the platforms that reflected some of the authenticity, like the documentations from OpenAI, papers published with esteemed publishers, and authentic interviews. Furthermore, the analytical figures, facts, and hypothesis were verified thoroughly by the team through various other resources in order to report in the study. For further clarifying the proposed work's current iteration aspect. The current iteration of the proposed work was updated on 25th March 2024.

## 3. Sustainability

Since November 2022, chatGPT has been a hot topic for researchers and consumer industry alike. A lot of studies either focus on the future of applications by integrating chatGPT to consumer electronic devices and its variants or how chatGPT can advance the LLMs in order to achieve artificial general intelligence (AGI). One of the least talked about issues concerning LLMs, specifically chatGPT is its sustainability in the context of greenhouse gases and carbon emissions. Greenhouse gases and carbon emissions contribute directly to climate change [13]. One of the ways to dive into sustainability issues related to chatGPT is the consideration of its environmental cost. In this article, we discuss the environmental cost with respect to the carbon footprint. The carbon footprint can be discussed for (a) training process, (b) inference, and (c) the complete life cycle [14]. We will discuss the carbon footprint from the perspective of training, inference and life cycle process, respectively. The carbon footprint for a machine learning model can be determined by the electricity consumption and its associated carbon intensity. Electricity consumption also undertakes the hardware employed while the carbon intensity is more deviated towards the way electricity is produced, i.e. wind energy, solar energy, coal or nuclear energy. Unless the exact details are known, the estimation can be done by computing average carbon intensity relative to the electricity grid location.

### 3.1 Sustainability for Training LLMs like ChatGPT

The LLMs are designed to generate accurate text based on the queries. The training of LLMs is carried out on large-scale datasets for potential usage in text generation, machine translation, and chatbots. The model during the training process is fed with lots of text to adjust the model weights. The process of training is considered to be computationally intensive, thus, the main reason for relating the process to carbon footprint. Most of the LLMs are based on transformer architecture that require vast amount of text data to be trained on. As mentioned earlier, the transformer networks use attention mechanism that extract positional embedding to find correlation among words that are semantically similar. The training process of LLMs requires the data to be presented in two categories, i.e. input and output. The former is the input query, and the latter is the one that needs to be predicted which represents the succession to the input query. The training for optimizing parameters is normally performed through standard neural network backpropagation algorithm. The chatGPT is built on GPT-3.5 which comprises of 355 billion parameters, suggesting that these parameters need to be adjusted or tuned to provide reasonably accurate results. A general assumption is that the training is performed only once, but the right set of parameters could not be found right away with just one go. Therefore, it is safe to assume that the network is trained multiple times until it yields satisfactory results. End users might just fine-tune the pre-trained network; however, it also requires multiple attempts to adjust the parameters and yield satisfactory results. Although the chatGPT is one of the most popular LLM but it's not the largest LLM, yet. So far, Google's PaLM and opensource BLOOM are larger with 500 billion and 176 billion parameters, respectively.

In order to understand why the carbon footprint is an important topic to discuss relative to GPT-3 and chatGPT, we need to understand the basic dynamics. The GPT-3 is trained on Common Crawl datasets, which, as of October 2022 had 3.15 billion pages that sums up to 418 Terabytes of data. Subsequently, GPT-3 needs to optimize 175 billion parameters on 418 Terabytes of data that might exhibit instability during the training process. A study [15] suggested that carbon intensity varies at different places of the world specifically based on the energy sources that are used to power up the grids. Nuclear power, hydro, and solar power generation sources yield the least amount of carbon intensity while oil, coal, and natural gas result in high-end carbon intensity. Over the past years, some studies have carried out analysis on the carbon footprint of LLMs [16–20]. For instance, Bannour et al. [20] performed carbon footprint analysis relative to LLMs using six different tools. One of the tools is publicly available that computes the carbon footprint of LLMs [21]. Narayanan et al. [22] assumed that the GPT-3 model required 34 days to train with 1024 A100 GPUs using 300 billion tokens and a batch size of 1536, respectively. 1024 A100 GPUs usage over 34 days roughly equals to the computation time of 835.5k hours. In order to determine the carboon footprint, we also need to consider the cloud provider or the region where the training was performed. As per the study [23], AWS Canada (Central), Azure Canada (East), and GCP Europe (West6) yield the lowest carbon footprint. The Canada uses hydroelectric power whereas the Switzerland operates on carbon neutrality initiative. These three regions are at the lower side of spectrum, while Azure South Africa (West) and Azure South Africa (North) are at the high-end of carbon footprint spectrum as they use oil and coal as their power generation sources. Considering the Occam's razor, we use the cleanest energy source and carbon intensive energy source for assuming the carbon footprint for training a GPT-3. We also provide a comparison with some of the carbon footprint sources to provide a reasoning behind the sustainability discussion in Table 1.

Table 1 Comparative analysis of average CO2 emissions/year with different sources and GPT-3

| Source | CO2 emissions per year (tons) |
|---|---|
| Boeing 747 (Heathrow to Edinburgh) 530 Kms [25] | 400 |
| Passenger Vehicle (11,500 miles/year) [26] | 4.6 |
| Average American [24] | 16 |
| Average Person (Non-American) [24] | 4-5 |
| Training GPT in least carbon intensive area (once) | 4 |
| Training GPT in most carbon intensive area (once) | 200 |

The above analysis is just an estimation, due to its simplistic approach and limited data released by their respective companies. However, recent studies conducted a carbon footprint analysis of GPT-3 and Meta's OPT training processes [17,19] and found that the latter emits 75 metric tons whereas the former emits 500 metric tons, which is 2.5x more than what we estimated. The above analysis is quite important as there are various competitors and various organizations that either train LLMs like GPT-3 or fine-tune them that require similar amount of CO2 emissions. Figure 1 only summarizes some of the LLMs that are available. Let’s estimate a reasonable number, i.e. 100 LLMs are trained for approximately 100 times (a reasonable guess), which results in around 10K training sessions. Just simply multiplying the carbon emissions (200 tons) with 10K training sessions would result in 2,000,000 tons of emissions. It should be noted that this only accounts for training as of now. With the popularity and progression of LLMs, the number would rise to a significant level, respectively. Furthermore, this number would also change if the number of parameters increased to trillions, which GPT-4 is assumed to be.

### 3.2 Sustainability for LLM Lifecycle and inferences like ChatGPT

Another aspect of sustainability in a model’s life cycle is the inference process. Recent article [23,24] highlighted the energy problem concerning Generative AI studies as foundational. The study suggests that MidJourney[2], a generative AI bot along with ChatGPT has redefined the term popularity in AI space in 2023. However, this popularity incurs a stupendous amount of energy cost for its realization and interaction with users. A recent study [25] suggested that the AI technology, specifically the generative AI and LLMs are predicted to consume around 29.3 terawatt-hours per year, which is equivalent to the energy consumed by Ireland (an entire country). Furthermore, the report highlights that training process of LLMs typically consumes 1,000 megawatt-hours of electricity and the energy consumption for inferences made by the users will be much higher than the consumption of energy incurred in the training process. Therefore, it is important to consider the LLM lifecycle and inference into consideration rather than only considering the training cost. An AI startup Hugging Faces proposed an efficient way of calculating carbon emissions in their recent paper [18]. It would be a better opportunity not only for AI tech companies but also for governments, regulators, and technology auditors to evaluate the environmental impact of such LLMs. The paper measures the carbon footprint of their own LLM BLOOM with respect to the training of the model on a supercomputer, electricity cost of manufacturing the hardware of a supercomputer, and inferential energy required to run BLOOM after its deployment. The carbon footprint of the inferential process was computed using CodeCarbon tool [26] that computed the carbon emission throughout the course of 18 days. The paper [27] estimated that the model inferential process yields 19 Kilograms of CO2/day, which is equivalent of driving a new car for 54 miles.

Recently, Facebook (Meta) also released the statistics of the carbon footprint analysis for their latest LLM LLaMA that outperformed GPT-3 on many language-oriented tasks [28]. Electricity consumption is provided for the whole model life cycle instead of only training process. A brief comparison of the electricity consumption for popular LLMs is shown in Table 2. Considering that GPT-3 has higher number of parameters than LLaMA combined, it is assumed that Facebook reported the electricity consumption for their failed attempts as well (5 months of training period) in comparison to GPT-3 that reported (14.8 days of training period). It is a good gesture from top tech giants to report the model life cycle energy consumption and we hope that it continues to provide a reality check when designing trillion parameters LLM.

---

[2] https://www.midjourney.com/home

Table 2 Carbon Footprint of LLMs in KWh and equivalent Danes (average power consumption of 1 Dane is 1600 KWh).

| LLM | Total Power Consumption | Equivalent to Danes |
|---|---|---|
| OPT-175B Meta [17] | 356,000 KWh | 222 |
| BLOOM-175B HuggingFaces [18] | 475,000 KWh | 297 |
| LLaMA-7B Meta [28] | 36,000 KWh | 22 |
| LLaMA-13B Meta [28] | 59,000 KWh | 37 |
| LLaMA-33B Meta [28] | 233,000 KWh | 146 |
| LLaMA-65B Meta [28] | 449,000 KWh | 281 |
| LLaMA-combined Meta [28] | 2,638,000 KWh | 1649 |
| GPT-3-175B OpenAI [29] | 1,287,000 KWh | 804 |

For the inferential part, a thorough computation of energy consumption for chatGPT's inference is given in [30]. One of the recent articles by Patel and Ahmed [31] assumes the number of active users for chatGPT to be 13 million and it was also assumed that 15 queries were made by each of the active users per day. Therefore, around 29k NVIDIA A100 GPUs would be required to serve chatGPT in its inference process. With the above information the multiplication of 13 million with 15 requests would yield around 195 million daily requests. Accumulating the requests for a month period would yield 5.85 billion requests, accordingly. In BLOOM's paper it was estimated that the BLOOM takes around 0.00396 KWh of energy to handle each request. Assuming that chatGPT takes the same amount of energy, it would amount to 23,166,000 KWh based on monthly requests, which is equivalent to 14,479 Danes, respectively. The above computation does not undertake the monthly energy of the GPU usage. NVIDIA A100's maximum power draw was computed to be 0.4 KW[3]. The aforementioned study by Patel and Ahmed assumes that the idle time should be factored in, therefore, the hardware should be assumed to be operating at 50% capacity. In this regard, the average power draw would be down to 0.2 KW. As per the estimates, chatGPT uses 29K GPUs, which would amount to 5,800 KW for an hour. Given the aforementioned assumed data, we can compute the GPU's monthly electricity consumption for chatGPT to be $30 x 24 x 5800 =$ 4,176,000 KWh, which is equivalent to 2,610 Danes, respectively. It should be noted that the computations are hypothetical based on the numbers provided in the aforementioned study. OpenAI does not provide the electricity consumption for GPT-3. Although the numbers are hypothetical and leverage the information from existing studies, the assumption still provides a consumption bracket that could be used as a basis to revise policies and regulations. There are several LLMs that facilitate real-time requests and use several GPUs. The numbers provided above would double and increase exponentially with the increasing number of conversational bot providers in coming years.

### 3.3 Estimated Training Cost of LLM

Estimated costs play an important part in impacting the affordability of carrying out research while sustainability of the environment. Over the years, LLMs have been scaled up in number of parameters, which is directly proportional to the training costs. For instance, in 2019 GPT2 was released that incurred an estimated cost of 50k USD to train. In comparison, 540 billion parameter model PaLM incurred an estimated cost of 8 million USD. From the above example it can be deduced that number of parameters play an important role in determining the estimated training cost. Recently, an AI Index report [32] carried out research where costs of multiple LLMs were estimated. However, the estimated costs are moderately reported, and we provide some references that the estimated costs are higher than the reported ones. In the report, the highest reported cost for training LLMs is for Megatron-Turing NLG 530 billion parameters amounting to 11.35 million dollars while the second and third spots are taken by Gopher and PaLM bearing

[3] https://www.nvidia.com/content/dam/en-zz/Solutions/Data-Center/a100/pdf/a100-80gb-datasheet-update-nvidia-us-1521051-r2-web.pdf

8.55 and 8.01 million dollars. On the far end, GPT-3 and BLOOM incur the cost of 1.80 and 2.29 million dollars, respectively.

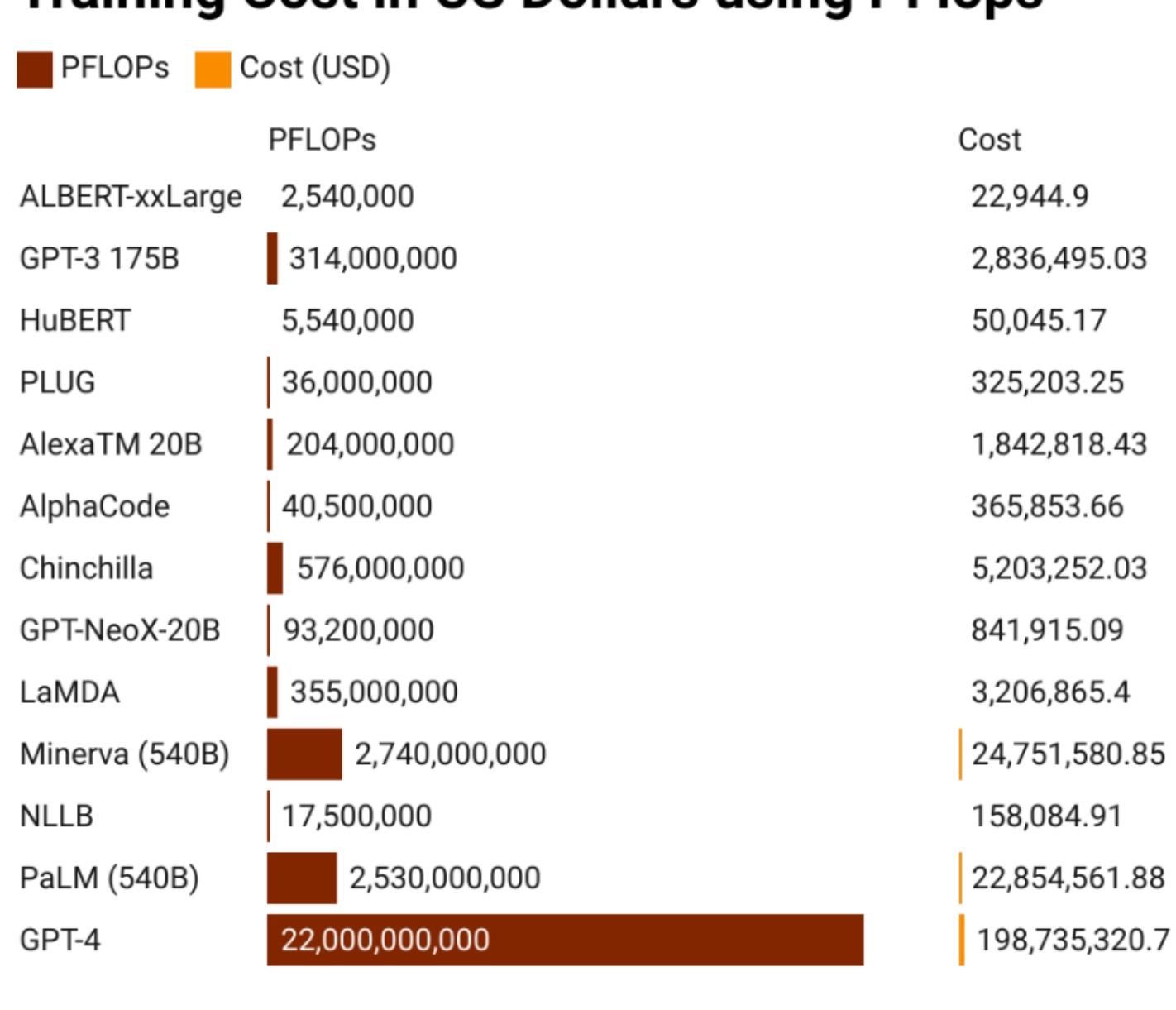

Figure 3 Estimated cost of different large language models using PetaFlops as Basis. The data is obtained from [33] and the assumption is based on the study [34]

Now let's base our results on some estimated and hypothetical facts. Most of the LLMs are computed on cloud, thus the amount to train LLMs can reach up to millions of dollars depending on the service provider. The paper [28] suggested that their model yielding 65 billion parameters took 21 days to train on 2048 GPUs having RAM of 80GB each. A popular choice that meets the aforementioned requirement is NVIDIA A100 GPU. According to Google Cloud Service Provider[4], the hourly price of NVIDIA A100 GPU is around 3.93$. Now the information provided in [27,35] the training cost can be estimated as 21 days x 24 hours x 3.93 USD x 2048 GPUs would yield 4.057 million dollars. Similarly, based on the study [23] GPT-3 requires 1024 A100 GPUs trained for 34 days; therefore, the computation can be carried out as 34 days x 24 hours x 3.93 USD x 1024 GPUs, which results in 3.284 million dollars. It should be noted that this is a one-time training estimated cost and studies have proved that the model is required to be trained multiple times in order to reach its optimal performance.

Another approach to compute the cost is with respect to the number of FLOPs [34]. One of the estimation methods that they use is:

- Renting a TPU instance would provide an estimate cost for floating point operations (FLOP)
- A similar approach can be applied to other cloud providers by extracting the cost per FLOP.

As per Google Cloud, it charges around 32 USD/hour for 32 cores TPUv3 pod. A study [36] suggested that 123 TFLOP/sec are provided by TPUv3 chip. While assuming that the peak utilization factor is around 50% [34] the study estimated 110.7 PFLOPs per dollar. We use the assumption and illustrate the training cost for various models using the data provided by [33] in Figure 3. The computation is performed on the models from 2020-2023 based on the data availability.

[4] https://cloud.google.com/compute/gpus-pricing

### 3.4 Mitigation and Recommendation

Reducing the carbon footprint of large language models is an important consideration for promoting sustainable and responsible AI development. Here are some ways in which language models can improve their training and inference process to minimize their environmental impact and reduce their carbon footprint:

- Optimize Compute Resources: Language models can be trained and run on energy-efficient computing resources, such as low-power CPUs, GPUs, or specialized hardware like Tensor Processing Units (TPUs). These energy-efficient hardware options can help reduce electricity consumption during training and inference, leading to lower carbon emissions.
- Use Renewable Energy Sources: Data centers and computing infrastructure that power language models can be powered by renewable energy sources, such as solar, wind, or hydroelectric power, to minimize the carbon footprint associated with electricity consumption. This can be achieved through partnerships with green data centers or investing in on-site renewable energy generation.
- Fine-tune Training Data: Fine-tuning, which is the process of training a pre-trained model on a smaller dataset, can help reduce the overall training time and computational resources required. By carefully curating the training data to include diverse and representative samples, models can achieve good performance with less data, thus reducing the environmental impact associated with large-scale data processing.
- Optimize Model Architecture: Improving the model architecture and algorithmic efficiency can reduce the computational requirements during training and inference, leading to lower energy consumption and carbon emissions. Techniques such as model pruning, quantization, and distillation can be employed to optimize model size, complexity, and computational requirements.
- Implement Dynamic Resource Allocation: Language models can dynamically allocate computational resources during training and inference based on the actual workload requirements. This can involve scaling up or down the resources based on the model's performance and accuracy requirements, thereby optimizing energy consumption and minimizing carbon emissions.
- Reduce Redundant Computation: Language models can avoid redundant computation during training and inference. Techniques such as caching, memorization, and incremental training can be employed to minimize redundant computation and reduce energy consumption.
- Encourage Collaboration: Collaboration among researchers and organizations can help share resources and expertise, leading to more efficient and sustainable AI development. Open-source initiatives, shared datasets, and collaborative research efforts can foster innovation while reducing the duplication of resources and efforts, thereby minimizing the environmental impact of language model development.
- Raise Awareness and Education: Raising awareness among researchers, developers, and users about the importance of environmental sustainability in AI development can lead to more conscious decision-making and practices. Education and training programs can help promote best practices for reducing the carbon footprint of language models, including energy-efficient computing, renewable energy usage, and model optimization techniques.

Reducing the carbon footprint of large language models involves a combination of optimizing compute resources, using renewable energy sources, fine-tuning training data, optimizing model architecture, implementing dynamic resource allocation, reducing redundant computation, encouraging collaboration, and raising awareness and education. By adopting these strategies, language models can contribute to a more sustainable and environmentally responsible AI development process.

## 4. LLM (ChatGPT) Privacy Concerns

The rise of LLMs is irresistible, [5]which is obvious when we look at subscription numbers. The chatGPT is one of the fastest platforms to have 100 million active users concerning consumer applications. Many startups have been launched that are built upon chatGPT. However, one of the issues that is not given enough attention concerning LLMs commercial usage is the privacy concern. A few days ago, Google released its conversational bot (Bard) that is only allowed to users above 18 years old, but it shows a pattern that tech companies are eager to launch their own conversational bots to mark their entry in the given space. One of the problems concerning privacy with the LLMs and their commercial usage is that they are fueled by personal data. A few articles shed some light upon the privacy issue while assuming that the data on which chatGPT is trained is systematically scraped from posts, websites, articles, books, and personal information without proper consent. Now one may ask why it is considered to be a privacy concern? The main reason is consent. It is probable that comments, product reviews or blog posts written by individuals have been consumed by chatGPT for training purposes. However, consent was not given to OpenAI for using the data, which is a privacy violation, especially if it is indicative of one's personal information or identification. Even the usage of publicly available data can cause a breach of contextual integrity [37] is considered to be a privacy violation, suggesting that the information might not be used in the same context as it was intended. Furthermore, OpenAI stores individual data such as personal information, which is partially in accordance with the General Data Protection Regulation (GDPR) and in some countries their compliance with GDPR is still questionable [38]. One of the examples is the recent ban of chatGPT in Italy over data breach involving payment information and user conversations on 20 March 2023. Also, the watchdog suggested that there is no means of verification for the users whether they are of an appropriate age to use chatGPT. Therefore, some responses generated by chatGPT might not be suitable for users belonging to underage group. Several European countries are also looking into it, for instance, Irish government. Certainly, this ban highlights the importance of compliance with regulatory bodies in order to protect individual's privacy information. A legislation process for AI-based systems has already been initiated in Europe but such an AI act would take years to take significant effect. It should also be noted that chatGPT has been blocked in other countries as well that include Russia, North Korea, Iran, and China.

[5] https://gdpr-info.eu/art-17-gdpr/

Apart from individual's privacy, some of the data which was used in the training process of chatGPT was copyrighted or proprietary. For instance, a snapshot from one of the queries we passed in chatGPT, i.e. Write an article on "Towards Industrial Private AI: A Two-Tier Framework for Data and Model Security”, is shown in Figure 4. Although the idea and the motivation of the text has been borrowed from [39] but it is neither cited nor given credit to the original article. This shows that the copyrighted text was consumed by OpenAI’s chatGPT. There has been a lot of debate on OpenAI’s approach to use the scraped data as the individuals whose data has been consumed by chatGPT were not compensated, however, the company's monetary worth has been doubled since 2021. Furthermore, OpenAI has also launched chatGPT plus[6] which is a subscription-based plan, and it is estimated to generate a revenue of around 1 billion dollars by the end of next year. In addition, as per chatGPT's privacy policy [40] it collects information such as user interaction data with the site, browser settings and its type, and IP address, along with the content type that users consider interacting with chatGPT. They also collect information concerning browsing activities across websites and over a certain period of time. Privacy policy also states that *"In addition, from time to time, we may analyze the general behavior and characteristics of users of our services and share aggregated information like general user statistics with third parties, publish such aggregated information or make such aggregated information generally available. We may collect aggregated information through the Services, through cookies, and through other means described in this privacy policy."* Another statement made by chatGPT's privacy policy states that *" In certain circumstances we may provide your Personal information to third parties without further notice to you, unless required by the law"*. Experts have been analyzing privacy concerns associated with chatGPT.

A very recent online article [41] also highlighted potential privacy concerns over chatGPT and some associated bugs that make conversation titles and chat histories of some users available to see. Although OpenAI CEO Altman accepted the glitch and stated that the issue has been resolved, it shows that the platform is not vulnerable to cyber-attacks or differential privacy attacks [39,42–45], thus the private information concerning users and their conversations can be potentially at risk.

Another issue that has been making rounds in recent times is the open availability of the LLM models such as ChatGPT, BLOOM, LLaMA, Bard, and others. Most of the LLMs operate in closed access rather than their open counterparts. For instance, Google Bard and OpenAI’s ChatGPT-4 are only accessible through their APIs to the developers, which, in turn, provides privacy guidelines. On the other hand, Meta’s LLaMA had been open source for a while with minimal protections that can ignite the spark in the hacker community to use LLaMA for spreading misinformation, harassment, spamming, fraud, privacy violations, and cybercrimes. Although it has nothing to do with the Meta's intention as they want to make the technology open-sourced, however, the recent report from CyberArk highlighted that such open models and technology without built-in ethical guidelines can be used for enhancing identity security threat landscapes.

---

[6] https://openai.com/blog/chatgpt-plus

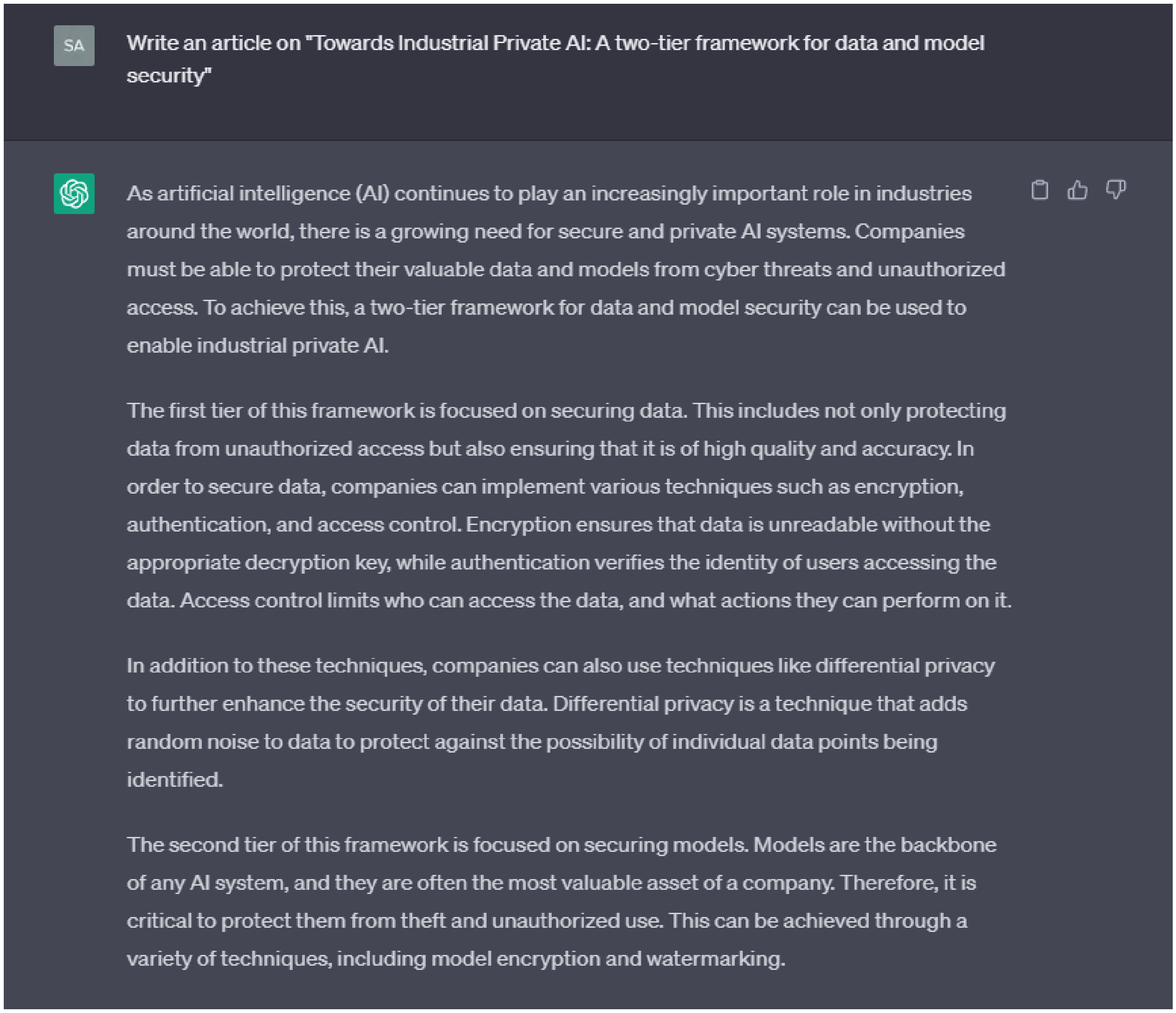

Figure 4 An example of chatGPT showing paraphrased copyrighted text.

The study also highlights a survey-based figure, which predicts that the organizations will have to face AI-powered attacks in coming years. The same claim was made by Fortune [46], which terms LLMs as force multiplier for privacy and security threats. The study suggested that the email-delivered attacks were increased in 2023 that used LLMs in one way or another. It also suggests that such email-delivered attacks accounted to be 86% of the overall attack shares. Considering the problems regarding openness, Meta AI has been considering limiting the access of LLaMA through their various iterations, especially the ones made in 2024 with Purple LLaMA [47]. Nevertheless, there is no doubt that LLMs have changed the privacy threat landscape and have the potential to disrupt the security of large organizations if not regulated in a proper manner.

### 4.1 Mitigation and Recommendation

Addressing privacy concerns is crucial for ensuring responsible use of large language models. Here are some ways in which language models can improve their policies and models to reduce privacy issues:

- Data Privacy Protection: Language models can implement strong data privacy protection measures, such as data anonymization, aggregation, and encryption, to prevent unauthorized access or misuse of user data during training and inference. User data should be handled with strict adherence to privacy regulations and best practices to minimize the risk of privacy breaches.
- Consent and Control: Language models can provide users with clear and transparent options to consent and control the collection, use, and storage of their data. This can include explicit consent mechanisms, privacy settings, and user-friendly interfaces that allow users to easily understand and manage their privacy preferences.
- Differential Privacy: Differential privacy is a privacy-preserving technique that adds noise or perturbation to the training data or model parameters to protect the privacy of individual users while maintaining the overall model's accuracy. Implementing differential privacy mechanisms can help prevent unauthorized inference or re-identification attacks and safeguard user privacy.
- Model Auditing and Explainability: Language models can implement auditing and explainability features that allow users to understand how their data is being used and provide insights into the model's decision-making process. This can enhance transparency and accountability and help identify and rectify potential privacy issues.
- Minimize Data Retention: Language models can minimize the retention of user data by only storing data necessary for the intended purpose and for the minimum duration required. Regular data purging and retention policies can be implemented to reduce the risk of data breaches and unauthorized access.
- Federated Learning: Federated learning is a distributed machine learning approach where the model is trained on local devices or servers, and only aggregated model updates are shared, instead of raw data. This can help protect user data by keeping it locally and reducing the need to share sensitive data with central servers.
- Robust Security Measures: Language models can implement robust security measures, such as encryption, authentication, and access controls, to protect against unauthorized access, data breaches, and other security threats. Regular security audits and updates can be conducted to ensure the model's security posture is maintained.
- Ethical Data Usage Policies: Language models can implement ethical data usage policies that clearly outline the principles and guidelines for data collection, use, and sharing. This can include avoiding biased or discriminatory data, respecting user preferences and privacy rights, and adhering to ethical and legal standards.
- User Education: Educating users about the privacy implications of large language models, their data usage policies, and the importance of protecting their privacy can empower them to make informed decisions and take necessary precautions while using the models.

Improving policies and models to address privacy concerns involves implementing data privacy protection measures, obtaining consent and providing user control, implementing differential privacy, enabling model auditing and explainability, minimizing data retention, adopting federated learning, implementing robust security measures, defining ethical data usage policies, and promoting user education. By implementing these measures, language models can mitigate privacy risks, accordingly.

### 5. Digital Divide

Since the launch of chatGPT, it has been clear that the platform can boost the creativity and productivity of students, teachers, researchers, content creators, and others. From the perspective of development, it is yet to be observed who will benefit the most from chatGPT, and how it will impact the low-income countries and workers in the Global South [48–50]. However, there is no denying that chatGPT is more affordable in comparison to human-like AI assistants such as Google Assistant, Alexa, and Siri as they require google devices, echo dot, and iPhone, whereas chatGPT requires internet access and basic literacy level. Technological changes over the years have shown that it creates both winners and losers. It is all about adjustment and adapting to technological changes to retain one's value. The workers that adapt will retain or get their value increased while the ones that don't will be obsolete and lose to the AI paradigm shift. On the brighter side, it creates new job spaces and develops a market for specific services and goods.

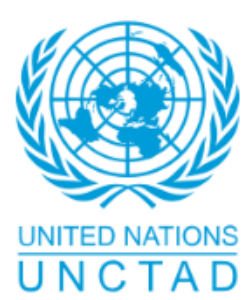

**Low-income countries are less prepared to benefit from AI**

Mean download internet speed versus share of skilled workers of total working population

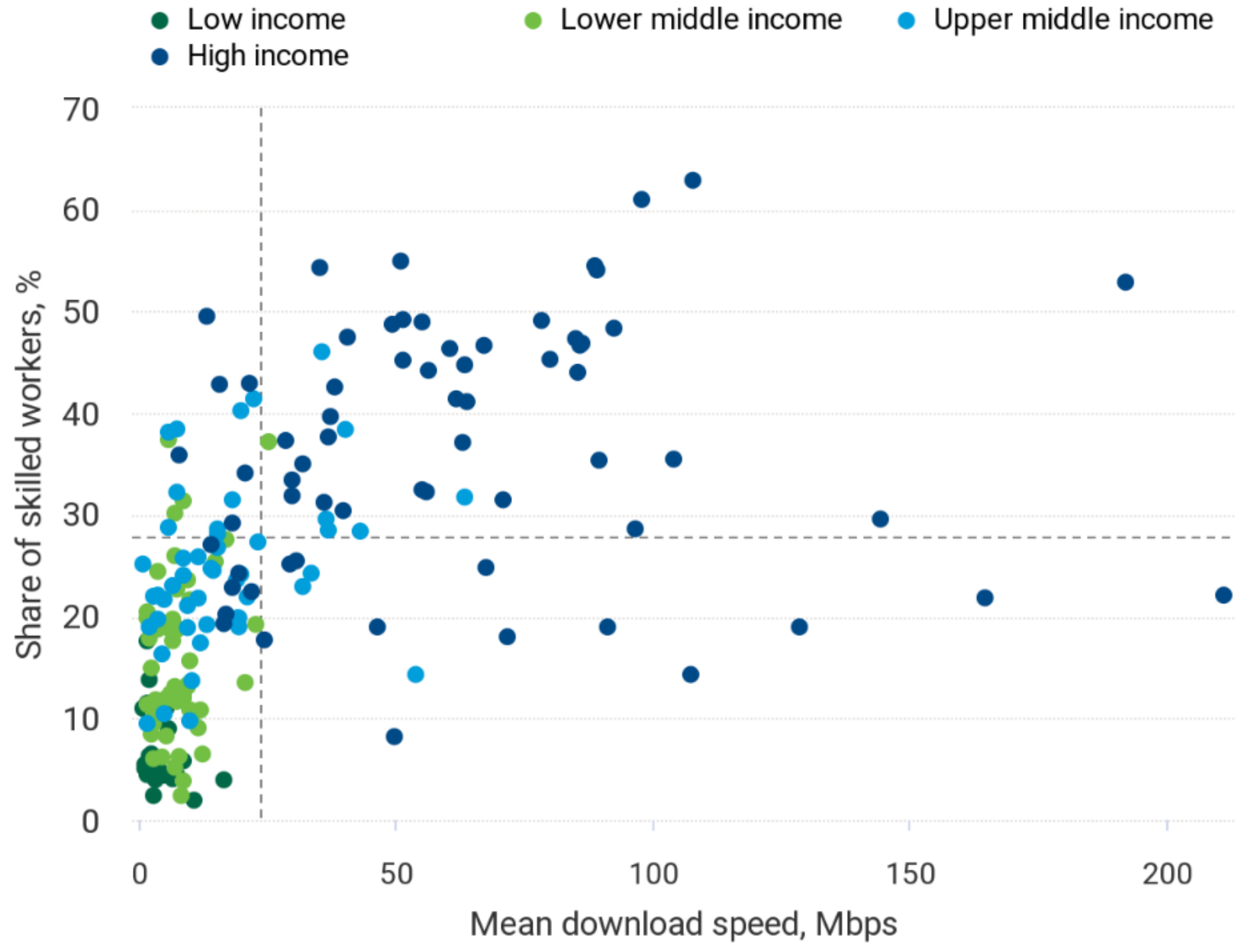

*Source:* UNCTAD based on data from ITU and M-Lab.
*Note:* Dotted lines represent averages

Figure 5 Skilled workers versus mean Internet speed for high income, upper middle income, lower middle income, and low-income countries. The dotted line in the graph refers to the average values. Graph courtesy: UNCTAD [48], and the data was collected by M-Lab and ITU.

Since COVID-19, there was a rise in the number of telemigrants[7] from developing countries that performed their jobs in the capacity of software developers, legal clerks, accountants, and X-ray analysts for firms in the developed countries. This gig allowed the workers from developing countries to compete with skilled workers across the globe, get experience in the concerned field, and get a reasonable monetary benefit. It has been predicted by the experts that the existence of LLMs such as chatGPT risks the jobs of telemigrants. It is also predicted that most workers and firms operating in developing countries will not be able to make the most from LLMs such as chatGPT due to the unavailability of high-speed internet and high-skilled labour, thus, creating a digital divide between high-income and low-income countries. A study from United Nations Conference on Trade and Development (UNCTAD) [48], provided the data regarding number of skilled workers and mean download speed (Mbps) and suggested that low-income or upper-low-income countries lag in the high-speed internet as well as share of skilled workers, therefore, they are slower in adoption of digital technologies. A visual illustration of their study is shown in Figure 5. The data suggests that lower income countries like Ethiopia, Guinea-Bissau, and others have a mean download speed of 1 Mbps and 5% skilled workers relative to the total working population. Burundi, Cabo Verde, and Burkina Faso have the least number of skilled workers, i.e. 2%, the mean Internet speed varies from 3 Mbps to 11 Mbps, respectively. Countries like Bangladesh and Pakistan have a mean Internet speed of 3 Mbps while the share of skilled workers vary from 9% to 10\%. India has a mean Internet speed of 23 Mbps with the share of skilled workers to be 19%. United states and China have mean Internet speeds of 92 Mbps and 2 Mbps with the share of skilled workers as 48% and 19%. We assume that the discrepancy is due to the population gap. However, the highest mean Internet speed is from

---

[7] https://unctad.org/meeting/cstd-side-event-public-lecture-professor-richard-baldwin

Liechtenstein with 211 Mbps while the highest share of skilled workers is with Luxembourg having 63% of its population skilled in some capacity. Min, Median, and Max values for average Internet speed and shared skilled workers are provided in Table 3 and 4.

It is quite evident from the data that there is a huge gap between low income and upper middle income in terms of average Internet speed, let alone be compared with high income countries. For instance, the maximum average Internet speed in low-income countries is 11 Mbps which is less than that of Median average Internet speed for upper middle-income countries and almost 6x less than that of maximum average Internet speed for upper middle-income countries. Furthermore, the maximum average Internet speed for high income countries is 19x and 8x more than that of the maximum average Internet speed for low and lower middle-income countries. Similar trends can be noted for the share of skilled workers as well. Another readiness study conducted by UNCTAD [48] also suggests that the developing countries generally encounter problems while adapting, adopting, and using frontier technologies such as research and development, digital infrastructure, and skills, in our case LLMs such as chatGPT.

Table 3 Min, Median, and Max for Average Internet Speed among Low, Lower middle, Upper middle-, and High-Income countries.

| Category | Country | Mean Internet Speed |
|---|---|---|
| **Minimum** | | |
| Low Income | Yemen | 1 Mbps |
| Low Income | South Sudan | 1 Mbps |
| Low Income | Ethiopia | 1 Mbps |
| Low Income | Guinea-Bissau | 1 Mbps |
| Low Income | Afghanistan | 1 Mbps |
| Lower Middle Income | Timor-Leste | 1 Mbps |
| Lower Middle Income | Djibouti | 1 Mbps |
| Upper Middle Income | Turkmenistan | 1 Mbps |
| Upper Middle Income | Equatorial Guinea | 1 Mbps |
| High Income | French Polynesia | 8 Mbps |
| **Median** | | |
| Low Income | Burundi | 3 Mbps |
| Low Income | Niger | 3 Mbps |
| Lower Middle Income | Uzbekistan | 7 Mbps |
| Lower Middle Income | Samoa | 7 Mbps |
| Lower Middle Income | Tunisia | 7 Mbps |
| Lower Middle Income | Bolivia | 7 Mbps |
| Lower Middle Income | Iran | 7 Mbps |
| Lower Middle Income | Honduras | 7 Mbps |
| Lower Middle Income | Senegal | 7 Mbps |
| Lower Middle Income | Kyrgyzstan | 7 Mbps |
| Lower Middle Income | Nepal | 7 Mbps |
| Upper Middle Income | Armenia | 18 Mbps |
| High Income | Solvenia | 67 Mbps |
| High Income | Romania | 67 Mbps |
| **Maximum** | | |
| Low Income | Burkina Faso | 11 Mbps |
| Lower Middle Income | Ukraine | 25 Mbps |
| Upper Middle Income | Bulgaria | 63 Mbps |
| High Income | Liechtenstein | 211 Mbps |

Aforementioned was the case from technological and development perspective, however, digital divide is also created amongst students due to the available Internet speed. Some experts from tech advocacy group [51–53] suggested that tools like chatGPT can help students remove writer's block on several tasks. Similarly, researchers and academicians have also suggested that students either not using such tools or do not have access to will be at disadvantage. However, the use of such tools is largely associated with basic knowledge and Internet speed, which enhances the digital inequality among the students from upper-high income countries and mid-lower income countries, respectively. Industrial Analytics Platform in conjunction with UNIDO conducted a study on chatGPT search trends in conjunction with human capital index [54–56] and showed that there is a positive correlation between the two. It should also be noticed in their study that the higher end of human capital index that searches for chatGPT is mostly from high-income countries that also supports our hypothesized concern. Furthermore, we leveraged the data from similarWeb[8] for the chatGPT webpage and depict the traffic share of top 50 countries in Figure 6. It was also shown that 61.05% of traffic share was from high-upper middle-income countries while 25.93% of the traffic share was from low-lower middle-income countries. Interesting facts can be observed that if India's share alone from low-lower middle-income countries is 10.67%, if taken out the category only has 15.93% of traffic share. In addition, the only low-income country listed in the top 50 is Nigeria with 1.10%. This supports our hypothesis of the digital divide created by chatGPT.

Table 4 Min, Median, and Max for Share of Skilled Workers among Low, Lower middle, Upper middle-, and High-Income countries.

| **Category** | **Country** | **Share of Skilled Workers** |
|---|---|---|
| **Minimum** | | |
| Low Income | Burundi | 2% |
| Low Income | Burkina Faso | 2% |
| Lower Middle Income | Cabo Verde | 2% |
| Upper Middle Income | Equatorial Guinea | 9% |
| High Income | Croatia | 8% |
| **Median** | | |
| Low Income | Gambia | 6% |
| Low Income | Sierra Leone | 6% |
| Lower Middle Income | Cameroon | 12% |
| Lower Middle Income | Papua New Guinea | 12% |
| Lower Middle Income | Angola | 12% |
| Lower Middle Income | Senegal | 12% |
| Lower Middle Income | Lesotho | 12% |
| Lower Middle Income | Honduras | 12% |
| Lower Middle Income | Cambodia | 12% |
| Upper Middle Income | Turkmenistan | 25% |
| Upper Middle Income | Georgia | 25% |
| Upper Middle Income | Saint Vincent and the Grenadines | 25% |
| High Income | New Caledonia | 35% |
| High Income | Spain | 35% |
| High Income | Hungary | 35% |
| **Maximum** | | |
| Low Income | Syrian Arab Republic | 18% |
| Lower Middle Income | Ukraine | 37% |
| Lower Middle Income | Lebanon | 37% |
| Upper Middle Income | Russian Federation | 46% |
| High Income | Luxembourg | 63% |

[8] https://similarweb.com

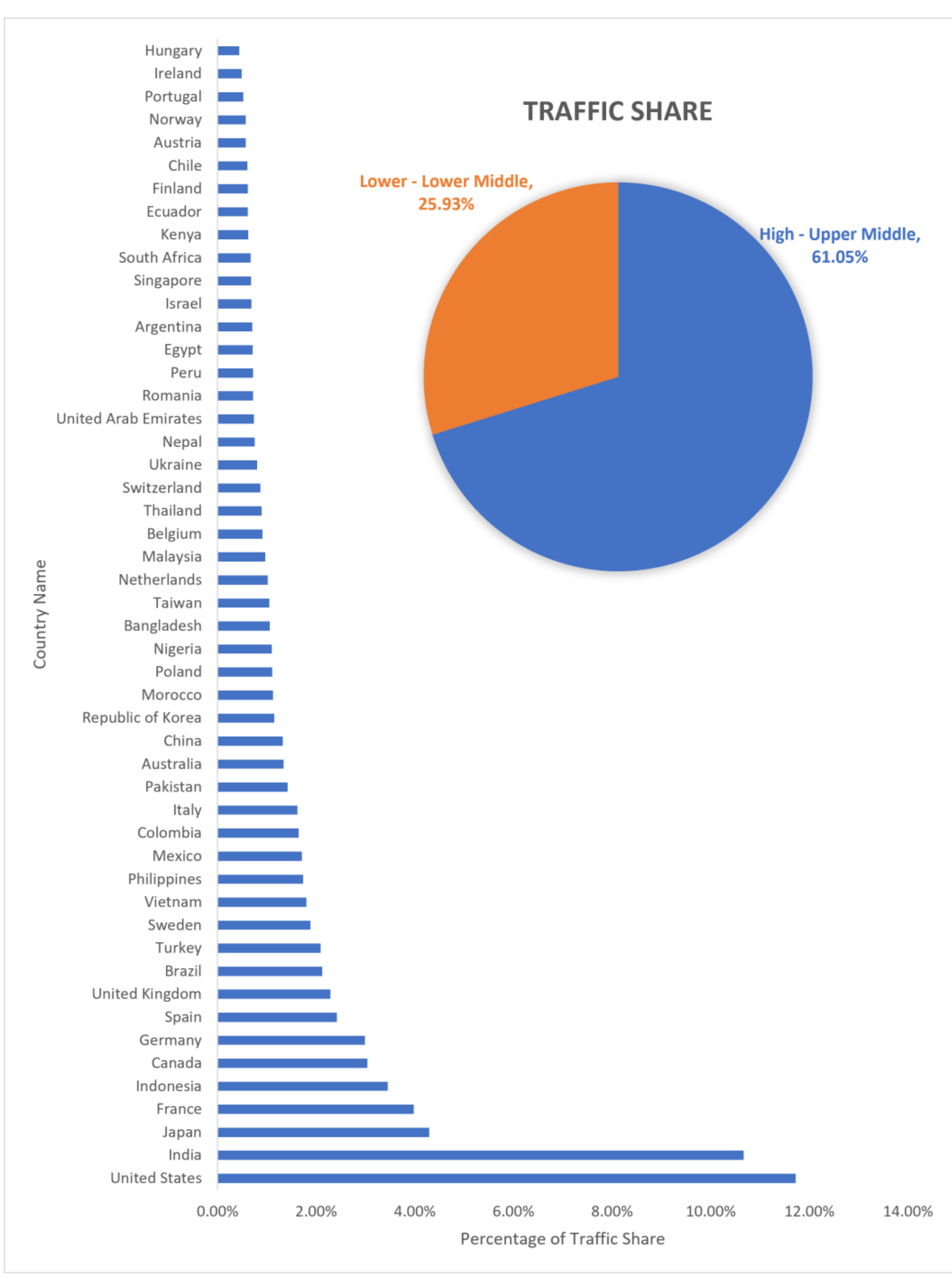

Figure 6 Traffic share of Top 50 countries for chatGPT website.

## 5.1 Mitigation and Recommendation

Reducing the digital divide, which refers to the gap in access to digital technologies and internet connectivity between different socio-economic groups and countries, is a critical challenge that large language models can help address. Here are some ways in which language models can improve their policies to close the digital divide gap between low-income, lower middle income, and high-income countries:

- Accessibility and Affordability: Language models can prioritize accessibility and affordability by offering free or low-cost access to their services in low-income and lower middle-income countries. This can include providing reduced data usage options, offering discounted or subsidized plans, or partnering with local organizations to make the services more affordable and accessible to users in these regions.

- Localization and Multilingual Support: Language models can improve their policies by prioritizing localization and multilingual support. This can include developing models that understand and generate content in local languages, dialects, and cultural nuances, making it more relevant and accessible to users in different regions. This can bridge the language barrier and enable users in low-income and lower middle-income countries to access and benefit from the services.
- Capacity Building and Training: Language models can contribute to closing the digital divide by offering capacity building and training programs to users in low-income and lower middle-income countries. This can include providing resources, tutorials, and training materials to help users develop skills in using the models for various applications, such as education, healthcare, and information retrieval. This can empower users in these countries to leverage the power of language models to address local challenges and improve their socio-economic opportunities.
- Partnerships with Local Organizations: Language models can collaborate with local organizations, such as non-profit organizations, academic institutions, and government agencies, to understand the specific needs and challenges of users in low-income and lower middle-income countries. This can help tailor the models' policies and offerings to better suit the local context and ensure that the benefits of the models are accessible and relevant to the target users.
- Infrastructure and Connectivity: Language models can work towards improving infrastructure and connectivity in low-income and lower middle-income countries. This can include partnering with internet service providers, telecommunication companies, and government agencies to improve internet access, connectivity, and infrastructure in underserved areas. This can enable users in these regions to have reliable and affordable access to the models' services.
- Social Responsibility and Ethical Considerations: Language models can prioritize social responsibility and ethical considerations in their policies. This can include adhering to ethical guidelines, avoiding biases and discrimination, and being transparent about data usage and privacy practices. By ensuring that the models are developed and used in a responsible and ethical manner, language models can build trust and promote inclusivity among users in different socio-economic settings.
- User Feedback and Iterative Improvements: Language models can actively seek feedback from users in low-income and lower middle-income countries and use it to drive iterative improvements in their policies and offerings. This can involve soliciting feedback through surveys, focus groups, or user testing, and incorporating the feedback into updates and enhancements to make the models more effective and user-friendly for users in these regions.

LLMs, specifically chatGPT can contribute to closing the digital divide gap between low-income, lower middle income, and high-income countries by prioritizing accessibility and affordability, localization and multilingual support, capacity building and training, partnerships with local organizations, infrastructure and connectivity improvements, social responsibility, and ethical considerations, and incorporating user feedback for iterative improvements. By taking these measures, language models can promote inclusivity and ensure that their benefits are accessible and relevant to users across different socio-economic settings.

## 6. Ethics

Large language models have raised various ethical issues related to privacy, bias, power, transparency, intellectual property, misinformation, and employment. This affects both the fairness of large language models as well as ethical concerns. Here are some of the main concerns:

- Privacy: Large language models require vast amounts of data to train. This data can include sensitive information about individuals, such as their conversations, search histories, and personal preferences. There is a risk that this data could be misused or accessed by unauthorized parties, which could have serious consequences for individuals' privacy.

- Bias: Large language models can inherit and amplify biases from their training data. For example, if a model is trained on a dataset that contains biased language, it may produce biased results. This could perpetuate and even worsen existing biases in society, such as racial, gender, and other forms of discrimination.
- Power: Large language models have the potential to shape and influence public discourse and decision-making processes. This gives the creators and users of these models a significant amount of power and responsibility. There is a risk that this power could be abused, intentionally or unintentionally, to manipulate public opinion or suppress dissent.
- Transparency: Large language models are often described as "black boxes" because it is difficult to understand how they arrive at their predictions or recommendations. This lack of transparency can make it difficult to identify and correct biases or other ethical issues that may arise.
- Intellectual property: There is also a debate about intellectual property rights related to large language models. Who owns the data used to train the models? Who owns the models themselves? These questions could have significant implications for the future of intellectual property law.
- Misinformation: Large language models can be used to generate fake news or misleading information, which can be spread rapidly through social media and other online platforms, leading to harmful consequences such as election interference or incitement of violence.
- Employment: Large language models can automate many tasks that were previously performed by humans, leading to concerns about job displacement and the need for retraining.

These are just a few examples of the ethical issues related to large language models. As this technology continues to develop, it is important to address these issues and ensure that it is used in ways that benefit society as a whole. In addition, there is a need for ethical frameworks and guidelines to ensure that large language models are developed and used in ways that are responsible and ethical.

Since the launch of chatGPT, researchers have tried to test the ethical boundaries of chatGPT. During its early release, a researcher from University of California, Berkeley's computation, and language lab shared snapshots of responses from chatGPT regarding a prompt asking, "Whether a person should be tortured". The response included some nationalities that chatGPT thought is OK to torture[9]. The article from The Intercept[10] shows several examples regarding the nationality of travelers that pose security risks to which chatGPT responded with the names of nationalities. Similarly, another example asked about the houses of worship that need to be put under surveillance and chatGPT responded with some anti-racial answers. Although, the article emphasizes that at first chatGPT is reluctant to provide specific answers (stern refusals) but with multiple tries (regenerate responses) it generates the aforementioned responses. Another web article shares an example of racial profiling that chatGPT was associated with can be accessed at[11]. As an AI language model, ChatGPT is programmed to generate responses to user inputs based on patterns and probabilities learned from vast amounts of data. However, there are potential ethical issues that may arise from its use. Here are a few examples:

- Bias: ChatGPT may exhibit biases in its responses if it has been trained on data that contains biases. For example, if the training data is biased against a particular group of people, ChatGPT may perpetuate these biases in its responses.
- Misinformation: ChatGPT may generate responses that contain inaccurate or false information, especially if it has not been trained on accurate and reliable sources of information. This can lead to harm if users rely on ChatGPT for advice or guidance.
- Privacy: ChatGPT may collect and store user data, including personal information, which could be used for unintended purposes, such as targeted advertising or surveillance.
- Responsibility: ChatGPT does not have a moral agency, and it cannot be held responsible for the consequences of its actions. However, those who create and deploy ChatGPT have a responsibility to ensure that it is used ethically and does not cause harm to users.

[9] https://twitter.com/spiantado/status/1599462405225881600
[10] https://theintercept.com/2022/12/08/openai-chatgpt-ai-bias-ethics/
[11] https://maktoobmedia.com/opinion/anti-reservationist-chatgpt-uncovering-racial-biases-in-ai-tools/

It is important to recognize these ethical issues and take steps to address them in order to ensure that ChatGPT is used in a responsible and ethical manner. ChatGPT, like other large language models, can strive to maintain fairness in its responses by using various techniques and approaches. Here are a few ways that ChatGPT may seek to promote fairness in its responses:

- Diversity and Inclusivity in Training Data: To minimize representation bias in its responses, ChatGPT's training data can be carefully curated to include diverse perspectives and underrepresented groups. This can help ensure that the model is exposed to a wide range of language patterns and language use cases that reflect the diversity of human communication.
- Counterfactual Data Augmentation: Counterfactual data augmentation is a technique used to help reduce bias in machine learning models. It involves artificially creating examples that counteract the biases present in the training data. By creating these counterfactual examples, ChatGPT can learn to recognize and mitigate biases that may be present in its training data.
- Debiasing Techniques: ChatGPT may also use debiasing techniques, which involve modifying the training data or modifying the model itself to reduce bias in the outputs. These techniques can range from simple modifications of training data to more complex algorithms that identify and remove biased patterns in the model's responses.
- Continuous Monitoring and Evaluation: ChatGPT can also continuously monitor and evaluate its responses to ensure that they are fair and inclusive. This can involve regularly testing the model's responses for bias or seeking feedback from users to identify and address areas where the model may be falling short.
- Guidelines for Human Reviewers: OpenAI provides guidelines to human reviewers who are involved in fine-tuning the model to avoid favoring any political, cultural, or religious group, including Muslims, and to refrain from generating Islamophobic or biased content. Reviewers are trained to be mindful of potential biases and to ensure that the model's responses do not promote hate speech, discrimination, or harmful stereotypes.
- User Feedback and Community Engagement: OpenAI encourages users to provide feedback on problematic outputs from ChatGPT and other models. This feedback helps identify potential biases and areas of improvement. OpenAI also actively engages with the research community, civil society, and other stakeholders to gather input and perspectives on mitigating biases and improving the fairness of its models.

It's important to note that achieving perfect fairness in machine learning models is a challenging and ongoing process, and there is always room for improvement. However, by employing these and other techniques, ChatGPT can work towards maintaining fairness and promoting diversity and inclusivity in its responses.

Over the time, users have provided feedback on problematic outputs from ChatGPT and other AI models. OpenAI actively encourages users to provide feedback on issues they encounter while using the models, including instances where the generated content may be biased, offensive, or inappropriate. This feedback is valuable in identifying and addressing potential biases and improving the model's behavior. However, based on general trends and common issues that can arise in content generation through chatGPT, some examples of outputs that could be marked problematic by users through their feedback. A recent report from Stanford, i.e. Artificial Intelligence Index Report 2023 [32] also highlighted ethical issues concerning chatGPT suggesting that it can be tricked into generating something that is not only unethical but also harmful on a macro society level. It is to be assumed that feedback would surely be provided to chatGPT on the mishap regarding the unethical behavior reported in [32]. Some examples of the feedback on unethical issues are listed below:

- Biased language: Instances where the model generates content that exhibits favoritism, prejudice, or discrimination towards certain groups of people based on characteristics such as race, gender, religion, or sexual orientation.
- Offensive or inappropriate content: Outputs that contain offensive, derogatory, or inappropriate language, including hate speech, profanity, or explicit content that may be considered offensive or objectionable.
- Misinformation or factual inaccuracies: Generated content that includes misinformation, false statements, or factual inaccuracies that could mislead or misinform users.

- Sensitive or controversial topics: Outputs that handle sensitive or controversial topics in a way that is insensitive, inappropriate, or biased, potentially perpetuating stereotypes or misconceptions.
- Incomplete or nonsensical responses: Outputs that are incomplete, nonsensical, or do not adequately address the user's query, resulting in an unsatisfactory response.
- It's important to note that user feedback is crucial in identifying and addressing these types of issues, and it helps in continuously improving the performance and behavior of AI models like ChatGPT.

Another ethical issue that is on the rise concerning chatGPT is related to the academics and students' integrity. Several states and universities have banned chatGPT so that the students could not cheat or plagiarize the content from chatGPT. According to the report [57] Tasmania, Queensland, and New South Wales, have banned chatGPT to promote novelty in their homework and assignments. Several other reports also address similar issues when it comes to academics [58–60]. Recently, at Boston university under the guidance of Wesley Wildman, prepared an initial draft that provides policies on the use of chatGPT and LLMs in academic settings. The policy was named as Generative AI Assistance (GAIA) policy[12].

### 6.1 Mitigation and Recommendation

Ethical concerns related to students copying assignments from large language models, such as ChatGPT, are important to address to ensure academic integrity and promote responsible use of the technology. Here are some ways in which LLMs can improve their policies to reduce ethical concerns related to students copying assignments and in general:

- Education and Awareness: Language models can prioritize education and awareness by clearly communicating to users, including students, the ethical considerations and responsible use of their services. This can include providing information on plagiarism, academic integrity, and the consequences of copying assignments from the model. Raising awareness among users about the ethical implications of copying assignments can help prevent unintentional misuse of the technology.
- Promoting Originality: Language models can promote originality in assignments by encouraging users to think critically, engage in independent research, and develop their own ideas and perspectives. This can be emphasized in the model's responses, suggestions, and prompts, which can emphasize the importance of original work and discourage direct copying from the model.
- Citation and Referencing: Language models can promote proper citation and referencing practices by encouraging users to acknowledge and attribute the sources of their information and ideas. This can include providing suggestions and guidelines on how to properly cite and reference sources in assignments, ensuring that users understand the importance of giving credit to original authors and avoiding plagiarism.
- Turnitin Integration: Language models can consider integrating with plagiarism detection tools, such as Turnitin, to enable users, including students, to check their assignments for potential plagiarism before submitting them. This can serve as a helpful tool for students to self-check their work and ensure that it meets the academic integrity standards of their institutions.
- Responsible Use Guidelines: Language models can provide clear and comprehensive guidelines for responsible use, including specific instructions on how the model should not be used for copying assignments or engaging in academic dishonesty. These guidelines can be prominently displayed on the model's user interface, website, or documentation, and should be easily accessible and understandable for all users, including students.
- Partnerships with Educational Institutions: Language models can collaborate with educational institutions, such as schools, colleges, and universities, to develop policies and guidelines that align with their academic integrity standards. This can involve consulting with educational experts, administrators, and faculty to understand their concerns and incorporate their feedback in the model's policies and offerings.
- User Authentication and Authorization: Language models can implement user authentication and authorization mechanisms to ensure that the model's services are accessed only by authorized users. This can involve verifying the identity and credentials of users, such as students, before granting them access to the model's services, and monitoring usage to detect and prevent misuse.

[12] https://www.bu.edu/files/2023/02/GAIA-Final-2023.pdf

- Continuous Monitoring and Improvement: Language models can implement continuous monitoring and improvement mechanisms to detect and address any potential misuse or ethical concerns related to students copying assignments. This can involve regular audits, feedback loops, and updates to the model's policies and guidelines to align with evolving ethical standards and best practices.

## 7. EU AI Act

### 7.1 Timelines and History

The initiative of the European union (EU) AI act was initiated first in October 2020, where the leaders focused on discussing the implications of AI and digital transitions. Initially, the idea was to provide clear distinction of high-risk systems that are associated with artificial intelligence, creating synergies, break communication barriers, and increase networking between the research centers and associated members, and lastly finding the ways for increasing investment towards AI through private and public organizations. While, the intention was clear, i.e. to increase the investment and develop more AI-based systems, the improvements in AI systems compelled the EU to diversify their thinking in their next meeting held in April 2021. The EU proposed its first AI act and a plan in order to regulate the use of AI in its member states. The focus of the EU AI Act was to improve trust in AI systems with the set of rules being laid out while fostering the update and development of AI technology. More details on the first proposal of EU AI Act can be found in [61]. The subsequent milestone was achieved on 6th December 2022, when a unified agreement was reached within the council to accept the general approach on AI act. The agreement was made in accordance with the fact that AI systems would respect EU values along with fundamental rights of its citizens and comply with the existing law. The proposal for adopting such a general approach was placed with the European parliament for further dialogues. Further details on the proposal are available at the press release from council of the EU, accordingly [62]. After months of negotiations and talks on the act, European parliament and the council agreed upon the provisional law on 9th December 2023. Permission was granted to formally prepare an AI Act text that safeguards the EU values and fundamental rights of the residents. Finally, on 13th March 2024, the plenary vote was obtained on EU AI Act's draft. Following the draft, the EU AI Act will follow the timelines as shown in Figure 7.

### 7.2 Summarizing EU AI Act

The EU AI Act is a landmark in the history of AI as it's the first legal and regulatory framework for AI-based systems and applications. The AI Act was proposed in order to make the AI systems and applications environmentally friendly, non-discriminatory, traceable, transparent and safe for usage in European Union states. The rationale for the act is that the regulation of AI should be done by humans rather than machines in order to avoid harmful outcomes. The main characteristic of EU AI Act is the levels that are defined to classify AI-based systems and evaluate risks associated with each level. The hypothetical infographic based on the information provided on EU AI Act has been depicted in Figure 8. Each of the levels are defined below:

- Low-Risk Systems: The low-risk AI systems are free to use, provided that they do not collect any personal information and do not violate the fundamental rights of the citizens. Furthermore, the low-risk systems assume that the AI system does not take any decision on behalf of the user. The systems that fall into this category are spam filters, and AI-enabled video games. In addition, many systems fall into this category that are in use today, provided that they do not violate the basic principle associated with low-risk systems.
-
- Limited-Risk Systems: The systems that are designed to hide transparency in their design and usage fall into the category of limited-risk systems. EU AI Act obliges the AI-system designers and companies to follow transparency regulations such that humans should be informed about the systems that are using are powered by AI. Therefore, it would be the user's choice to trust the decisions made by users. For instance, when using chatbots the users should be made aware that the interaction is being made by the machine rather than human so that they can make the informed decision of trusting the machine or not. An additional responsibility to the service provider is also obligated such that the content generated by AI is identifiable. Furthermore, the AI generated text that is available in the public domain should be labelled as artificially generated to inform the public in an explicit manner. The same rules apply to different forms of content including audio, image, and video modalities. It is assumed that such regulation would help to cope with the deep-fake systems.

-

- High-Risk Systems: High-risk systems are those that can affect the fundamental rights and safety of the citizens and EU member states. The high-risk systems are categorized into two, i.e. systems that reside under the EU's product safety legislation [61,63] and the systems that reside under areas require registration in EU database. The former includes but is not limited to lifts, medical devices, cars, aviation, and toys. The latter include legal interpretation assistance, law-related applications, border control management, migration and asylum control management, law enforcement, essential public and private services, worker and employment management, vocational and educational training, and operation and management of critical infrastructure.
  Such high-risk systems will be evaluated first before they are launched on the market and will also be monitored during their development cycle. Meanwhile, the right to complain to national authorities can be exercised by general public.

- Unacceptable Systems: These types of systems will be banned as they are considered to be a threat to the people. These systems include remote or real-time biometric systems, which include facial recognition, categorization of people through biometric identification, classification of people through their personal characteristics, socio-economic status and behavior, and products that invoke dangerous behaviors or manipulate the behavior of general public and specific vulnerable groups such as voice activated toys that invoke violent behavior in children.

There is also an excluded category, which we have not included in the infographic. The excluded category includes the models and systems that are designed for the sole purpose of scientific development and research. More exclusions are available, but they are provided to the government agencies and law enforcement agencies only.

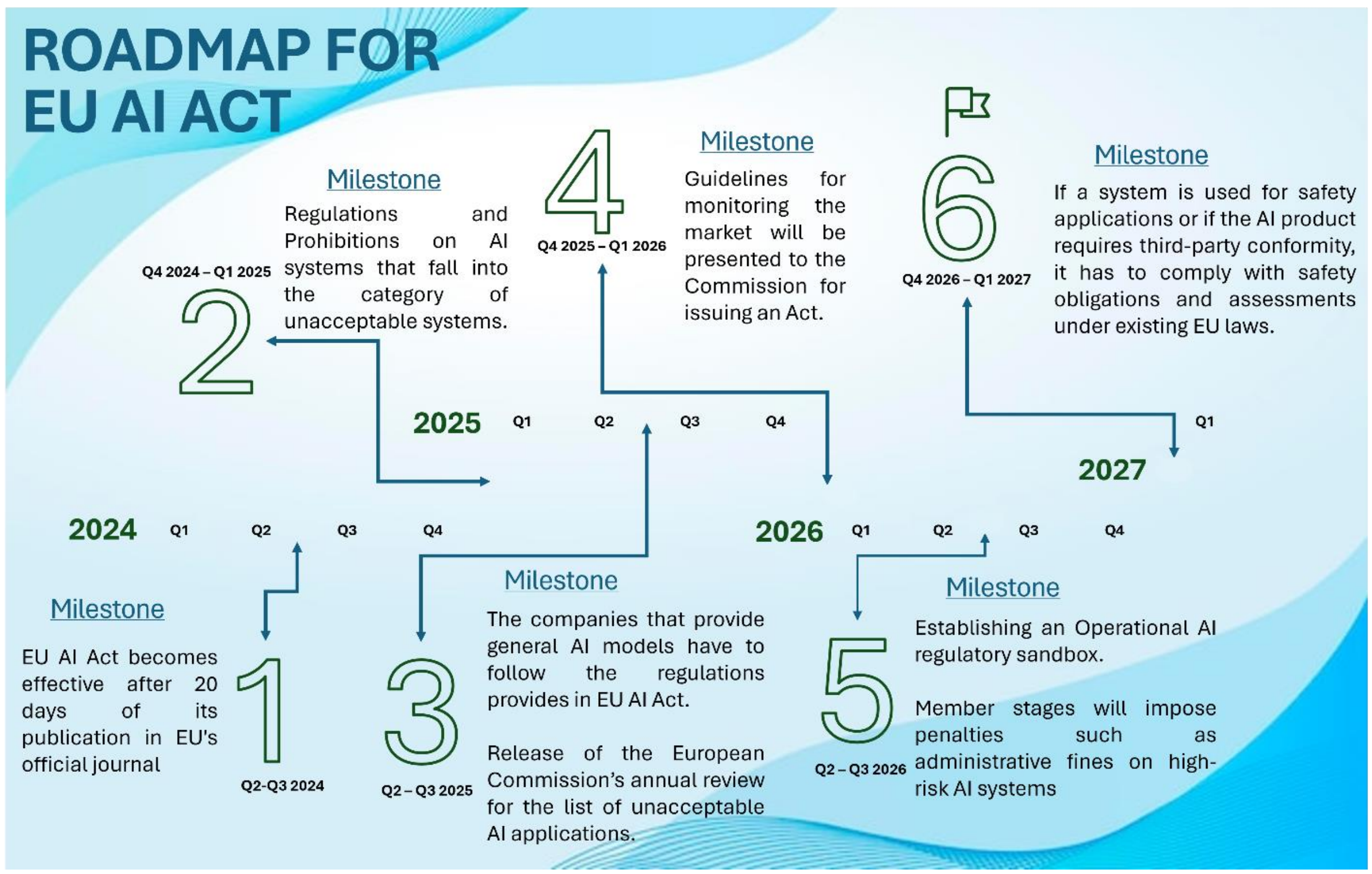

Figure 7 EU AI Act Implementation Plan with Timelines [63].

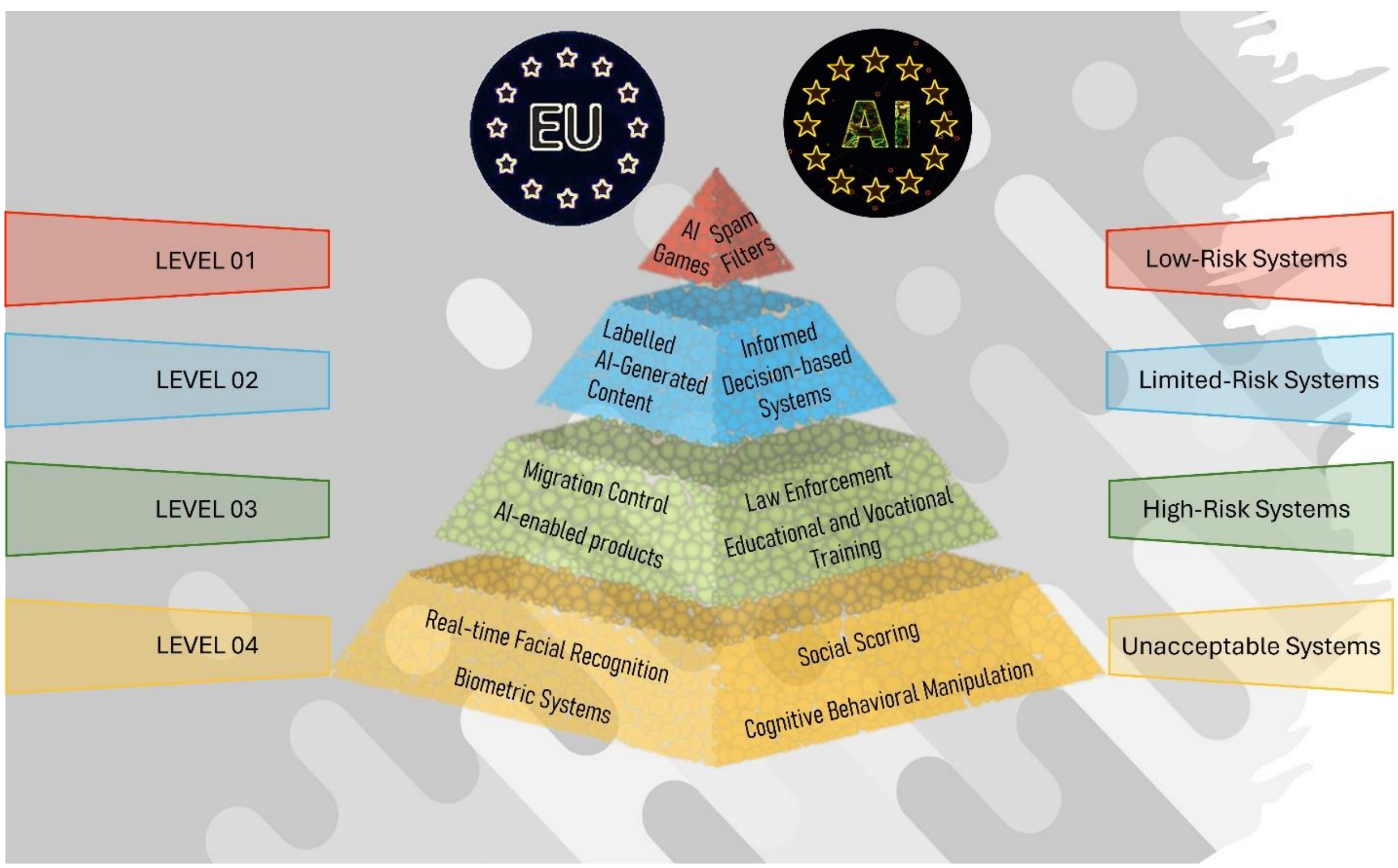

Figure 8 A hypothetical infographic that summarizes EU AI Act [63].

## 8. Lessons Learned

Although we have discussed extensively the SPADE evaluation that needs to be carried out for LLMs such as ChatGPT specifically, there is much more that needs more discussion, such as the type of energy sources being used for systems deploying LLMs or general-purpose AI. The systems need to undergo transparency and responsible AI checks to ensure the privacy and security of users. We have discussed the EU AI Act briefly; however, no such information is provided as to how such an act can deal with digital divide concerning the AI-based systems such as LLMs. Much emphasis has been made to ethical use of AI, however, the funding that gets available to train and deploy large LLMs. These are still not clear with respect to the EU AI Act. In this section, we suggest some policy recommendations that should be included if an AI policy act is constituted, specifically related to LLMs.

- Transparency: The policy should mandate transparency in the development, deployment, and operation of AI systems. This includes transparency in the data used for training, the algorithms used, and the decision-making processes of AI systems.
- Fairness and Bias Mitigation: The policy should require measures to ensure that AI systems are designed and implemented in a way that is fair and unbiased, without perpetuating discrimination or bias against any particular group or individual. This includes addressing issues such as bias in data, algorithmic bias, and unintended discriminatory impacts.
- Privacy and Data Protection: The policy should include provisions to protect the privacy and data rights of individuals, including guidelines for the collection, storage, and use of data in AI systems. This includes ensuring that AI systems comply with applicable data protection laws and regulations, and that data used for training and inference is handled securely and responsibly.
- Accountability and Liability: The policy should establish clear lines of accountability and liability for the actions and outcomes of AI systems. This includes defining responsibilities for developers, operators, and users of AI systems, and specifying the legal and ethical implications of AI-related decisions and actions.

- Human Oversight and Control: The policy should emphasize the importance of human oversight and control in AI systems. This includes ensuring that humans remain in control of decisions made by AI systems, and that AI is used as a tool to augment human decision-making, rather than replace it.
- Safety and Security: The policy should require measures to ensure the safety and security of AI systems, including robust testing, validation, and monitoring procedures. This includes addressing potential risks such as adversarial attacks, system failures, and unintended consequences of AI technologies.
- Ethical Considerations: The policy should highlight the importance of ethical considerations in the development and use of AI systems. This includes promoting transparency, fairness, accountability, and respect for human rights in all AI-related activities.
- Education and Awareness: The policy should include provisions for education and awareness programs to ensure that stakeholders, including users, developers, operators, and policymakers, are knowledgeable about the ethical, legal, and social implications of AI technologies.
- Stakeholder Engagement: The policy should mandate meaningful stakeholder engagement in the development, implementation, and evaluation of AI systems. This includes involving diverse stakeholders, such as affected communities, civil society organizations, and experts, in the decision-making processes related to AI technologies.
- Regular Evaluation and Update: The policy should require regular evaluation and update of AI systems to ensure compliance with the policy and adapt to changing technological, ethical, and societal considerations. This includes periodic review of the impact of AI systems on various dimensions, such as fairness, privacy, and human rights.

These recommendations are not exhaustive and may vary depending on the specific context and requirements of a statutory body. However, they provide a broad framework for policies that can help guide the development, deployment, and use of AI systems in a responsible, ethical, and accountable manner. It is crucial to involve various stakeholders in the process of formulating AI policies, including experts, policymakers, affected communities, and civil society organizations, to ensure a well-informed and inclusive approach to AI governance.

## 9. Conclusion

With the popularity of ChatGPT, LLaMA and its ongoing integrations, it is evident that there is a whole market space for large language models (LLMs). However, there are concerns that need to be addressed or policies need to be designed before the LLM market takes over. In this paper, we discuss several concerns related to chatGPT, specifically related to sustainability, privacy, digital divide, and ethics. Our hypothesized analysis shows that chatGPT consumes a lot of energy during both the training and the inferential phase. Such carbon footprint, if extended to various LLMS like chatGPT, we definitely be harmful for environment and would affect significantly to the climate change. We also show that there are several privacy concerns related to chatGPT, specifically how the data for the training was collected and how the data of individual uses is and will be used by OpenAI. With preliminary analysis, we also show that chatGPT is creating a digital divide among the low - lower middle income and upper middle - high income countries. Lastly, we show examples of concerns over ethics and fair usage of chatGPT.

The study also discusses the EU AI Act and provides mitigations and recommendations for each of the concerns in detail. Furthermore, we also provide suggestions for policies for AI policy act, if such policy is designed and presented on the governmental platform. We intend to improve this article over time by adding more details, updating the already provided details, and adding more preliminary analysis to support the facts.

**Authors Contribution Statement**

Sunder Ali Khowaja wrote the Sustainability, Digital Divide Part and designed infographics for EU AI Act timeline and summarization in the manuscript and compiled all the materials from other authors to prepare the manuscript.

Parus Khuwaja and Kapal Dev contributed to the Privacy part of the manuscript and helped in proof reading of the manuscript.

Weizheng Wang and Lewis Nkenyereye contributed to the Ethics part of the manuscript and helped in proof reading of the manuscript.

**Funding**

No funding was obtained for this study.

**Conflict of Interest**

The authors have no competing interests as defined by Springer, or other interests that might be perceived to influence the results and/or discussion reported in this paper.

**Data Availability Statement**

This manuscript does not report data generation or analysis.